# GAMA: towards a physical unde

Simon P Driver, Peder Norberg, Ivan K Baldry, Steven P Bamford, Andrew M Hopkins, Jochen Liske, Jon Loveday, John A Peacock and the GAMA Survey Team (Galaxy and Mass Assembly) review progress on the latest large galaxy redshift survey now underway on the 3.9 m Anglo-Australian Telescope.

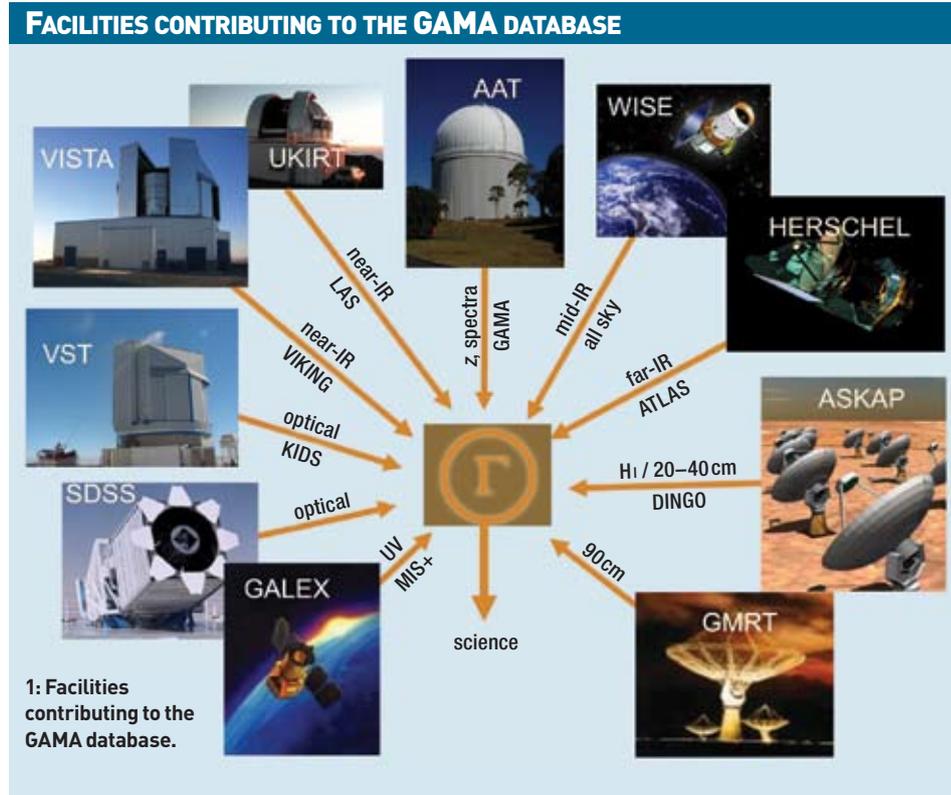

**Facilities contributing to the GAMA database**

1: Facilities contributing to the GAMA database.

Large international surveys of the low-redshift universe, such as the 2-degree Field Galaxy Redshift Survey (2dFGRS: Colless *et al.* 2001, 2003) and the Sloan Digital Sky Survey (SDSS: York *et al.* 2000; SDSSDR6: Adelman-McCarthy *et al.* 2008) have transformed our quantitative understanding of galaxies, galaxy populations and large-scale structure. This in turn has contributed directly to the emergence of a concordance cosmological model (e.g. Spergel *et al.* 2003, 2007; Cole *et al.* 2005): a flat, dark-energy dominated collisionless cold dark matter model (Lambda-CDM) which provides both the bedrock and blueprint for our galaxy formation models. Known as hierarchical-CDM, this structure-formation model provides a physically motivated, fully numerical and verifiable description of the observable universe on Mpc (and greater) scales. However, on the scales of clusters, groups and galaxies, the picture is less clear. There are significant inconsistencies between the basic mechanism (the hierarchical merging of dark matter haloes) and the empirical evidence (in particular the complex yet fragile sub-structure of individual galaxies); exactly how a merger-based process can give rise to such fine sub-structure lies at the heart of the debate. It is on these scales (between 1 kpc and 1 Mpc) that the poorly understood interplay between the dark and baryonic matter becomes crucial as the dark matter haloes virialize and merge, and the baryons decouple (from the dark matter) and crystallize into observable galaxies. The complexity of the physics and the range of length, time and mass scales involved (from atomic to cosmological) prevents full numerical modelling at this time. As a consequence, this regime can now be investigated only by phenomenological modelling which, by definition, is guided and informed by empirical datasets (e.g. Baugh 2006).

Insight into the physics underpinning galaxy formation is therefore being driven by the interaction between ever larger numerical simulations and technologically advanced observations leading to evermore sophisticated empirical datasets. The GAMA survey is designed to provide the best possible wide-area dataset for low to intermediate redshift galaxies that current technology allows (with some facilities still under construction, see figure 1). In particular the GAMA project will provide major steps forward in three key areas:

● **improved spectroscopic efficiency**, enabling the comprehensive sampling of lower-mass galaxies locally and higher-mass systems to intermediate redshifts while sampling the full range of environments, all within a single survey;

● **improved spatial resolution**, enabling the deconstruction of nearby galaxies into their distinct structural components (e.g. nuclei, bulges, bars and discs), believed to be relic imprints of their individual formation histories (e.g. AGN activity, merger, secular and accretion processes);

● **increased wavelength coverage**, enabling each galaxy's entire baryon (stars, dust and gas) and radiative energy budget to be quantified.

To achieve these advancements requires the coordination of an unprecedented spectroscopic and multiwavelength imaging programme incorporating major time allocations on ground-based facilities spanning four continents and three space missions (figure 1).

### Probing the CDM paradigm
At the heart of the GAMA project is the AAT spectroscopic survey, which was specifically designed to confront the popular CDM paradigm in three ways:

● measure the dark matter halo mass function (HMF) and its evolution using galaxy group catalogues probing down to Local Group sized systems;

● measure the baryonic systematics of galaxy formation: in particular the global Galaxy Stellar Mass Function (GSMF), star-formation efficiency and feedback;

● infer the galaxy merger rates over 5 Gyr via the observed number of close pairs and via structurally asymmetric systems.

The first of these tests addresses a direct prediction of numerical simulations, the second

# rstanding of galaxy formation

### Table 1: Facilities contributing to GAMA database

| survey | PI | wavelength | time on GAMA | obs. period |
|---|---|---|---|---|
| AAT-GAMA | S Driver | optical spectra | 66+99‡ nights | 2008–2012 |
| UKIRT-LAS | S Warren | near-IR imaging | 28 nights | 2008–2009 |
| VISTA-VIKING | W Sutherland | near-IR imaging | 60 nights | 2009–2010 |
| HERSCHEL-ATLAS | S Eales | far-IR imaging | 150 hours | 2009–2010 |
| GALEX-GAMA/MIS | R Tuffs/C Popescu | UV imaging | 50 hours | 2009 |
| WISE-All Sky | E Wright | mid-IR imaging | 100 hours (public) | 2009–2010 |
| GMRT-HATLAS | M Jarvis | radio continuum | 110 hours | 2009–2010 |
| VST-KIDS | K Kuijken | optical imaging | 96 nights | 2009–2010 |
| ASKAP-DINGO | M Meyer | radio line | 1 year | 2012+ |
| all | | | | 2012–2015 |

All time is guaranteed unless otherwise indicated.
‡Requires time allocation beyond mid-2010.

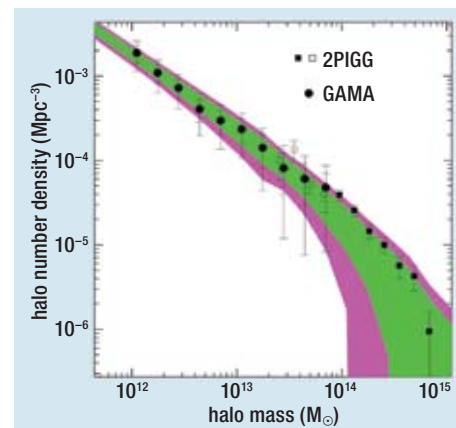

2: The dark matter halo mass function (HMF) along with predictions from the Millennium Simulation for a 90 degree$^2$ survey (pink shading) and a 250 degree$^2$ survey (green shading). Overlaid is the measurement of the HMF from the 2dFGRS 2PIGG catalogue (boxes) and our expected constraint from the final GAMA survey (solid symbols).

engages with the phenomenological assumptions and the third probes the basic mechanism for galaxy assembly in this paradigm. We now describe these issues in a little more detail.

### Measuring the halo mass function

Cold dark matter appears to have a mass density about 5.5 times that of normal matter, and is commonly assumed to be a relic particle from the very early history of the universe, when the typical photon energy was >100 Gev. Such a hypothesis might be verified by detecting γ-rays from CDM particle–antiparticle annihilations, but the interaction cross-section is generally so low that the CDM is often taken to interact via gravity only. A CDM particle whose abundance freezes out when it is highly non-relativistic will have extremely low random motions, so that small-scale density fluctuations survive. Because of the lack of direct interactions, these CDM structures grow efficiently under gravity early on, before matter–radiation decoupling. The baryonic material is subsequently affected by the gravitational force of the collapsing dark matter, and is drawn into the dark matter cores, dissipating its internal energy through collisional radiation and through angular momentum transfer with the emerging dark matter halo. Thus, although the anisotropies in the Cosmic Microwave Background (CMB) are dominated by 100 Mpc structures that subtend 1° today, the seeds of galaxy-scale structures already existed at the redshift of around 1100 when the CMB was generated. The key success of the CDM paradigm lies in this ability to account for very large-scale inhomogeneities, as seen for example by WMAP (Spergel *et al.* 2007), at the same time as providing understanding in principle of the origin of galaxies (e.g. Baugh *et al.* 2006).

When sufficient gas densities are reached, stars start to form (Kennicutt 1998), generating a self-gravitating CDM halo that contains gas and stars – i.e. a galaxy. In this picture, galaxies will act as test particles tracing the underlying CDM distribution, like lighthouses marking where the rocks are. But because of the hierarchical build-up of structure in the CDM paradigm, the correspondence between galaxies and CDM haloes is not one-to-one. When haloes merge into a new larger halo, the cores of the original haloes (together with the galaxies they host) survive for a while as subhaloes, but tidal stripping tends to diminish these as distinct mass concentrations – leaving just their baryonic cores. The limit of this process is a halo that contains a number of galaxies orbiting in a common potential, i.e. a group of galaxies. Galaxy groups thus serve as the clearest markers of CDM haloes, and the halo mass can be estimated from the relative orbital velocities of the galaxies in the group. Empirically, astronomers thus have to consider the halo occupation distribution: the number of galaxies hosted by a halo as a function of its mass and other properties (e.g. Peacock and Smith 2000, Berlind and Weinberg 2002). One of the main aims of GAMA is to measure this function.

Our primary science objective is therefore to assemble a large catalogue of galaxy groups, and to measure their velocity dispersions as a proxy for halo mass. This requires a survey that covers a large contiguous volume with high completeness in the sampling of the galaxy distribution as well as accurate redshifts. The first science product from such a catalogue will be the halo mass function, and the expected precision from different survey areas is shown in figure 2, along with current constraints from the 2dFGRS survey (2PIGG: Eke *et al.* 2004). The shaded regions show the expected range of halo mass functions drawn from the Millennium Simulation (Springel *et al.* 2005) for a random sample of 90 degree$^2$ regions surveyed to a depth of $r_{AB} < 19.4$ mag (pink; current survey status) and a random sample of 250 degree$^2$ regions surveyed to a depth of $r_{AB} < 19.8$ mag (green; final survey status).

The above description glosses over a number of important subtleties, which include the uncertainty in group membership; the efficiency of group-finding algorithms; and the relation between galaxy group velocity dispersion and

## A: From a galaxy distribution to a dark matter halo mass

To obtain an estimate of the underlying dark matter halo mass function from a galaxy redshift survey is far from trivial: very-low-mass groups contain too few galaxies to be detected robustly, and even massive clusters can have their properties biased by galaxies that are not assigned to the correct group. The means of combating these problems is to use realistic mock galaxy catalogues in order to measure and correct any biases in our analysis methods, and to optimize those methods. Our starting point is the GAMA galaxy distribution itself. Each galaxy comes with basic spatial properties (e.g. position on the sky and apparent distance based on its redshift, $z$) plus some intrinsic properties (e.g. flux $r_{AB}$, $K_{AB}$). By selecting a galaxy sample for which the survey selection function can be characterized with high accuracy, the observed galaxy distribution can be dissected into a collection of galaxy groups. The fundamental motivation here is the presumption from the CDM paradigm that all galaxies have dark matter haloes, and that each galaxy group corresponds to a unique parent dark matter halo.

There is a vast literature on the different methods for finding groups from a discrete distribution of objects. To test CDM it is vital that we subject the real and mock data to the same group-finding algorithm. In numerical simulations, dark matter haloes are most commonly defined using a standard friends-of-friends (FoF) algorithm, with one free parameter: the linking length $b$, whose value is usually ~20% of the mean (dark matter) particle separation. Once the dark matter haloes are defined, all galaxies within them are defined as being bound to that halo.

At the heart of the problem lie two fundamental differences between dark matter haloes and galaxy groups:
● the former can be roughly described as a continuous field, while the galaxies form a discrete distribution, even if surveys such as GAMA now provide a galaxy mean number density at least three times that of earlier surveys at similar redshifts;
● the dark matter density field in simulations

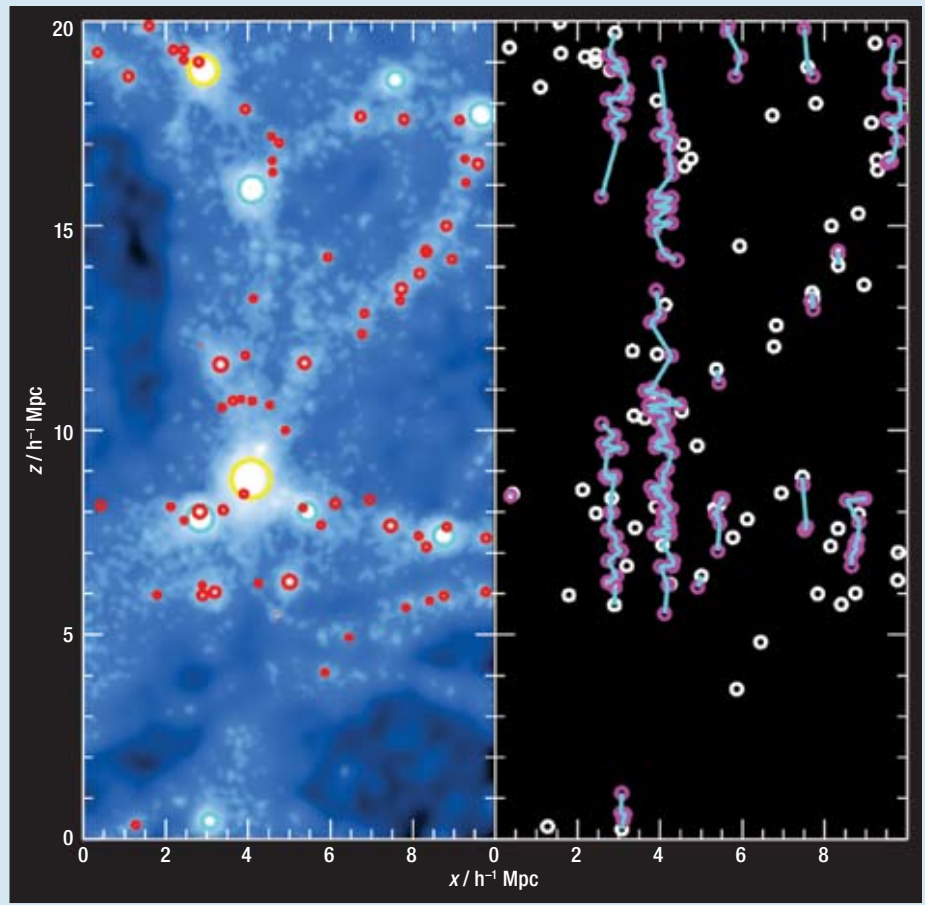

**3 (Left):** 4 Mpc thick slices through the Millennium Simulation showing the dark matter particles and bound groups. The group mass range is indicated by the colour circles (yellow $10^{14}$–$10^{13}$ $M_\odot$; cyan $10^{13}$–$10^{12}$ $M_\odot$; red <$10^{12}$ $M_\odot$). **(Right):** The same volume as on the left but giving the galaxy distribution in redshift space (i.e. with the additional velocity components due to the group velocity dispersion added). Dominant group galaxies are shown in white and other group members in magenta and connected with cyan lines. The figure illustrates the complexity involved in recovering the group associations and masses from the galaxy distribution observed in redshift space.

is accurately known in 3D, while the galaxy density field is only precisely known in 2D (i.e. in projection on the sky), with the third dimension affected by random galaxy velocities, leading to radial smearing of several Mpc, at least five times the average group size.

These two issues are highlighted in figure 3, which makes it clear that the route from a galaxy distribution in redshift space to a dark matter halo distribution requires several levels of tuning.

Once a galaxy group catalogue has been constructed, the dynamical mass estimate is straightforward and obtained by using a velocity dispersion estimator on the group members, together with an adopted scaling relation between velocity dispersion and halo mass. Like the fine tuning of the group catalogue, the relations between galaxy group velocity dispersion and dynamical mass are calibrated from realistic mock galaxy catalogues, for which we know both quantities.

---

underlying halo mass. Some of these, in particular the complexity of group finding, are discussed further in box A. In practice, these issues are dealt with by generating mock catalogues and simulating the observing process by applying the true spatial completeness maps and then using the same group-finding algorithms that are applied to the real data. The mocks shown in figure 2 have been instrumental in defining our required minimum survey area as 250 degree² with a required survey depth of $r_{AB,Limit}$ = 19.8 mag.

### Measuring the baryonic systematics of galaxy formation

Even from existing data on galaxies and groups, it is clear that the distribution of baryonic mass (starlight) does not directly follow the underlying dark matter. Understanding the physical processes that cause this difference is a fundamental challenge in galaxy formation. Considerable theoretical progress has been made (e.g. Baugh 2006), but the problem is sufficiently complex that further observational constraints are essential in order to achieve a robust understanding of the key baryonic processes. In particular, one of GAMA's key objectives will be to

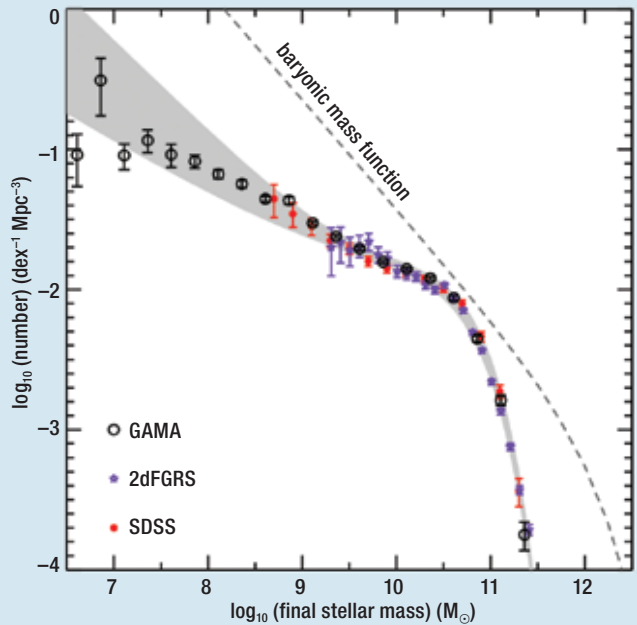

4: The Galaxy Stellar Mass Function (GSMF) from various recent studies (2dFGRS and SDSS), including preliminary results from the first two years of GAMA observations (open circles). The grey shaded region shows the approximate error range from the current GAMA data. Also shown as a dashed line is the expected galaxy mass function from numerical simulation coupled with a basic halo occupation distribution model. These two distributions combined fully constrain the star-formation efficiency.

### Table 2: Coordinates of initial three GAMA fields

| field | RA (J2000 deg.) | dec. (J2000 deg.) | area (deg.$^2$) |
|---|---|---|---|
| G09 | 129.0…141.0 | +3.0…−1.0 | 12×4 |
| G12 | 174.0…186.0 | +2.0…−2.0 | 12×4 |
| G15 | 211.5…223.5 | +2.0…−2.0 | 12×4 |

is therefore an important indication that, within the CDM paradigm, star formation cannot be equally efficient for all halo masses. This seems plausible (Baldry *et al.* 2008), as lower mass haloes will struggle to retain their baryons as they are heated and accelerated by SN winds during episodes of massive star formation. One important outcome of GAMA will be precise measurements of the extent of such "feedback" down to very low halo masses.

Figure 4 shows our preliminary measurement of the GSMF from GAMA, which extends significantly deeper than earlier studies and illustrates that the discrepancy seen at high mass continues to current detection limits. The results suggest that ~1% or less of the combined progenitor baryon mass in low-mass haloes is converted into stars. This raises the question of what and where are the remaining 99% of the baryons originally associated with low-mass systems? We plan to answer this question directly though deep radio observations with the Australian Square Kilometer Array Pathfinder (ASKAP: Johnston *et al.* 2007), which will measure the neutral H I and dynamical masses, as well as identifying any strong neutral gas inflows/outflows for most of our sample.

These preliminary results will be refined as we extend our understanding of completeness limits and selection biases in the GAMA dataset. For now, figure 4 simply emphasizes the potential for extremely accurate measurements of the star-formation efficiency. Beyond this, we intend to determine the dependence of efficiency not only on mass, but on morphology, environment and redshift. For this science, high completeness in the spectroscopic survey is essential, coupled with deep multicolour imaging to overcome surface brightness bias, and enable reliable stellar mass measurements inclusive of dust attenuation corrections.

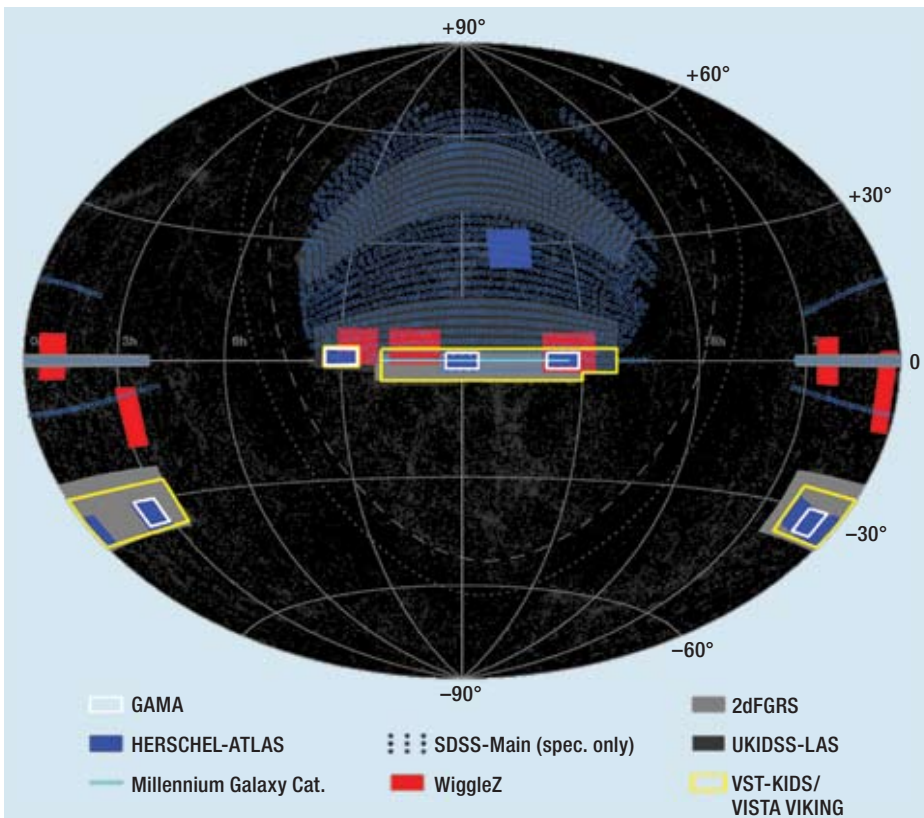

5: An Aitoff projection of the sky in equatorial coordinates centred at 12h, 0° showing the location of key surveys which are currently in progress or about to commence. The five GAMA regions are shown as white rectangles which overlap a significant portion of the Herschel-Atlas survey (blue solid regions).

measure the baryon fraction and star-formation efficiency as a function of DM halo mass.

To illustrate this connection more clearly, figure 4 shows the prediction of the galaxy baryonic mass function (dashed line) from Shankar *et al.* (2006). This is an attempt to predict the distribution of masses for dark-matter haloes that host single galaxies (i.e. excluding massive cluster-scale haloes), which is then scaled by the global ratio between the densities of baryons and dark matter. This simple baryonic mass distribution is compared with recent observations of the galaxy stellar mass function (GSMF; data points and shading). If star formation were equally efficient at all masses, these functions would mirror each other with a single baryonic mass-to-light ratio: this is clearly not the case. The gross discrepancy between the two distributions

### Quantifying merger rates

In our final test of the CDM paradigm, we explore the core mechanism underpinning galaxy assembly. In the standard picture, the hierarchical nature of density fluctuations leads to massive DM haloes being built via the merging of smaller ones. Because most if not all haloes are expected to contain galaxies, the halo merger rate as predicted by CDM must match, modulo some galaxy merging timescale, the fraction of galaxies in close physical pairs (indicative of imminent mergers: Patton *et al.* 2002) and the incidence of highly asymmetric or disturbed galaxies (indicative of ongoing or recent mergers: Conselice *et al.* 2008). The greater the fraction of galaxies in pairs and incidence of asymmetry, the greater the amount of merging. These effects should scale both with environment and lookback time, enabling a relatively sophisticated comparison with the halo merger trees within numerical simulations. The caveat in this test is the fact that the dynamical merger and asymmetry timescales for galaxies cannot be directly observed and must instead be obtained from simulations. Fortunately, the predictions for these timescales from high-resolution simulations of specific galaxy merger scenarios (e.g. Khochfar and Burkert 2006) are now quite robust and suggest that both timescales are of order 0.5–2 Gyr. This implies that several thousand close physical pairs and highly disturbed galaxies should be seen in the GAMA survey. Observationally, it is imperative to obtain high signal-to-noise sub-arcsec imaging data, and high-completeness spectroscopy of close pairs, in order to be able to identify disturbed galaxies and close physical pairs, respectively.

### The GAMA spectroscopic survey

To achieve our headline science goals of measuring the halo mass function and the efficiency with which galaxies form in different environments, the GAMA spectroscopic survey will eventually need to cover ~250 degree$^2$. In 2007 GAMA received an allocation of 66 nights of AAT time (over three years, taking GAMA up to the UK's withdrawal from the AAO in mid-2010) in order to begin the survey by tackling three equatorial fields of 48 degree$^2$ each (see table 2 and figure 5). These regions were selected to overlap with the planned VST KIDS and VISTA VIKING ESO Public Surveys and later adjusted to maximize overlap with the Herschel–ATLAS survey. The present time allocation will enable us to reach complete redshift coverage to limiting depths of $r_{AB} < 19.4$ mag in G09 and G15, and $r_{AB} < 19.8$ mag in G12, as well as $K_{AB} < 17.6$ mag over all three fields. However, to reach a uniform limiting depth of $r_{AB} < 19.8$ and $K_{AB} < 17.6$ mag over the full 250 degree$^2$ we need will require a further 99 nights.

At these final depths, GAMA will fully sample a source density of ~1150 degree$^{-2}$, which compares to densities of ~140 degree$^{-2}$ for the 2dFGRS and ~90 degree$^{-2}$ for the SDSS. Baldry *et al.* (2009) present a complete description of the GAMA input catalogues, which are based on data from SDSS DR6 (Adelman-McCarthy *et al.* 2008) and UKIDSS-LAS DR4 (Lawrence *et al.* 2007) while Robotham *et al.* (2009) describes the tiling strategy.

The spectroscopic survey is powered by the

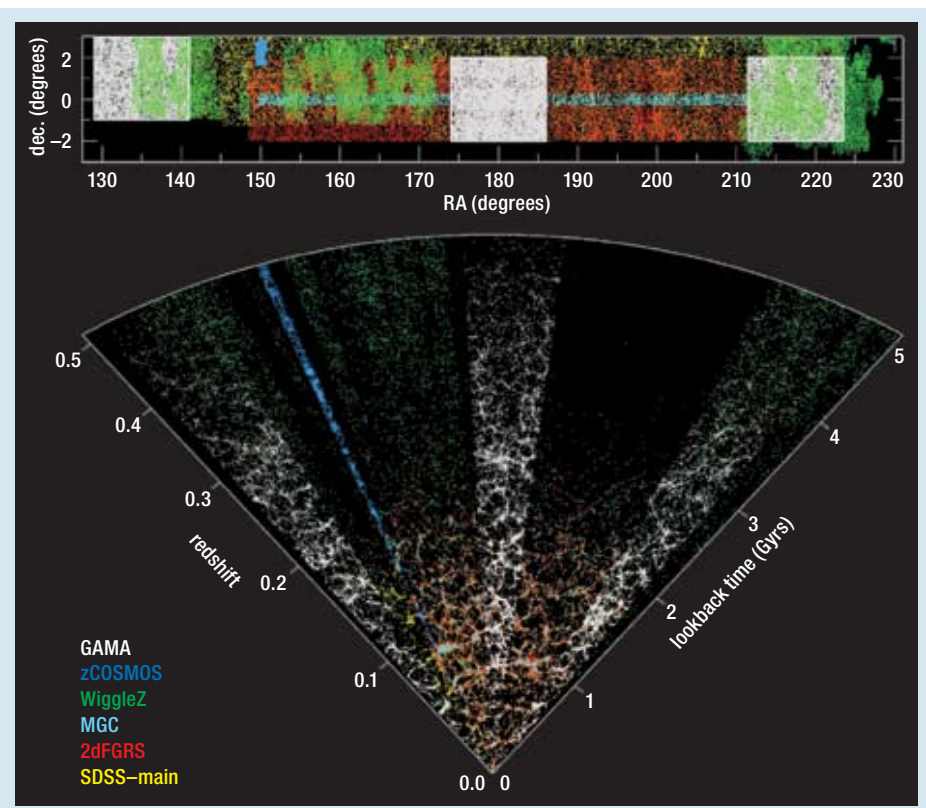

**6:** A lookback time/redshift cone plot for a 2° wedge of sky (–1°<δ<+1°) highlighting the current state of the GAMA survey (white data points). Also shown are data from the SDSS, 2dFGRS, MGC, zCOSMOS and WiggleZ surveys for this region of sky (shown in the upper panel). GAMA is the only study which can reveal the evolution of the large structure since *z*=0.5. Note that zCOSMOS has been relocated to lie within the 2° wedge.

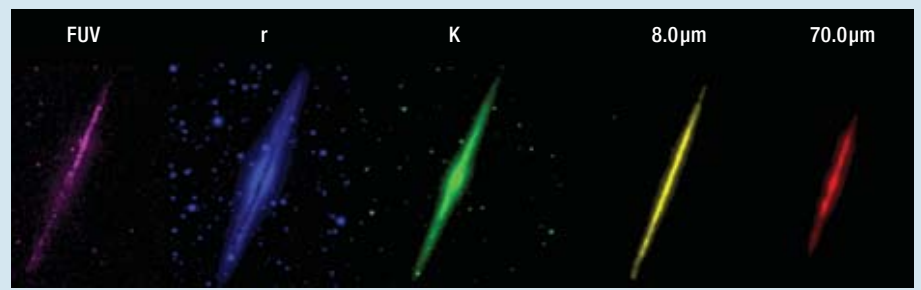

**7:** NGC891 as observed in far-UV, optical (*r*), near-IR (*K*), mid-IR (8.0μm) and far-IR (70μm) wavelengths. One can see significant variations in the image as one moves from filters sensitive to active star formation (FUV), young stars (optical), old stars (near-IR), warm dust (mid-IR) and cold dust (far-IR). All data were downloaded from the NASA Extragalactic Database or provided by the PI.

recently commissioned AAOmega spectrograph system (an upgrade of the original 2dF spectrograph: Sharp *et al.* 2006) which is used to obtain 5 Å resolution spectra from 3700 Å to 8800 Å for all galaxy targets inside the GAMA regions. The first two years of observation have been extremely successful, delivering over 90 000 new galaxy spectra. Figure 6 shows our current redshift cone plot, highlighting GAMA's comprehensive sampling of the large-scale structure out to *z*=0.5. Also plotted are the currently leading wide-area shallow surveys (SDSS, 2dFGRS, MGC), and one of the on-going pencil beam surveys (zCOSMOS). This comparison effectively illustrates GAMA's niche embodied by the comprehensive mapping of galaxy structures over a 5 Gyr lookback time: the shallow surveys have insufficient depth, while the pencil beam surveys offer insufficient cross-section to unveil the largest structures at low to intermediate redshifts. GAMA is presently the only intermediate redshift survey in this regime. When complete it will have acquired ~275 000 spectra; as of July 2009, the survey is about one third of the way to this final target.

### The broader multiwavelength GAMA survey

As stated in the opening section, GAMA is far more than simply a galaxy redshift survey: it comprises quality spectra, from which detailed line analysis work can be conducted, as well

## B: The GAMA database

Currently we envisage that the broader GAMA data will be processed by four distinct pipelines.

● **Structural analysis.** Based on GALFIT3 (Peng *et al*. 2002) all galaxies with $z<0.1$ will be spatially modelled in 2D to determine the intensity and shape parameters of any central nucleus, pressure-supported bulge, pseudo-bulge, bar, inner disc and outer (truncated) disc components in all optical/near-IR filters. For those systems with $z>0.1$, a single elliptical Sérsic profile will be measured (Graham and Driver 2005), enough to separate concentrated and diffuse systems and measure ellipticities and inclinations. Central supermassive black hole masses can be predicted from both the bulge Sérsic index and bulge luminosity in optical and near-IR wavebands (e.g. Vika *et al*. 2009). We will also measure concentration, asymmetry and smoothness (Conselice 2003) along with Gini coefficient and M20 measurements to enable classification by the ZEST+ code developed by the zCOSMOS team.

● **Spectral analysis.** All spectra will be flux calibrated using in-house software and then processed by a modified version of GANDALF (Sarzi *et al*. 2006, Schawinski *et al*. 2007) to separate the continuum (with its stellar absorption lines) from the nebular emission lines. Measurements of the emission-line properties will provide information on the ionization state of the gas, and hence the star-formation rate, AGN activity, gas metallicities and kinematics. Absorption-line measurements will provide information on the stellar population properties and hence on the star-formation histories, ages, metallicities, element abundances and velocity dispersion. See figure 8 for target lines.

● **Spectral energy modelling.** For those systems with detections in the far-IR, the total spectral energy distribution (SED) from UV to far-IR will be used to derive the intrinsic distributions of stellar emissivity and dust on an object-by-object basis using self-consistent SED model of Popescu *et al*. (2000). The model incorporates three distinct dust components (a thin disc, a thick disc and a clumpy component). This will provide measurements of dust opacity, temperature, mass, star formation rate and star formation history. For those systems without far-IR coverage, corrections for the optical light can be estimated using statistical methods calibrated on the far-IR sample, taking into account similar morphology and environment.

**All (~250k)**
**General:** GAMA ID, SDSS ID, $z$ (heliocentric), $z$ quality
**Flux:** UV, optical, near-IR, mid-IR, far-IR, radio (20, rest-21, 30, 40, 90cm)
**Shape:** CAS, Sersic index, half-light radii, b/a, PA in *ugrizYJHK*
**Opacity:** $\tau_{UV,ugriz,YJHK}$
**Spectral features:** Emission: H$\alpha$, H$\beta$, H$\gamma$, H$\delta$, O$_{II}$, O$_{III}$, N$_{II}$
Absorption: Dn4000, Ca4227, H$\alpha$, H$\beta$, H$\gamma$, H$\delta$, Mgb, Fe
**SFR:** UV, H$\alpha$, far-IR, radio continuum
**Fossil record:** age, SFH, element abundance
**AGN:** BPT diagnostics, type, strength, ionization state
**Dynamics:** $\sigma_{spec}$ (GANDALF), $W_{21}$, H$_I$ line profile
**Distances:** Tully-Fisher, Faber-Jackson
**Masses:** stellar, SMBH, H$_I$, dust, baryon, dynamical
**Environment/halo:** local density, group membership, group halo mass

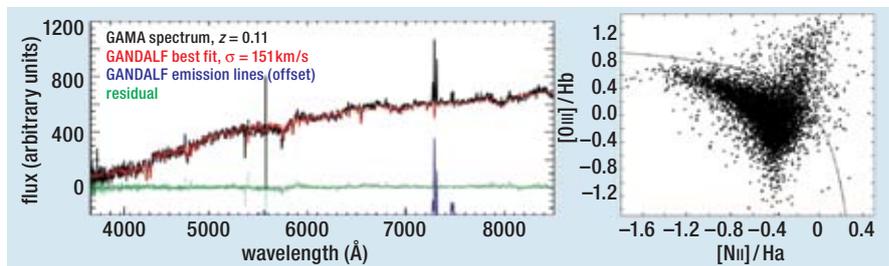

**For $z<0.1$ (~30k)**
**Structural:** bulge/bar/disc decomposition in *ugrizYJHK* (GALFIT3)
**Bulge:** Sersic index, half light radius, PA, ellipticity
**Bar:** Sersic index, half light radius, scale-length
**Disc:** scale-length, PA, b/a
**SMBH mass:** via M-$\sigma$, M-L, M-n relations

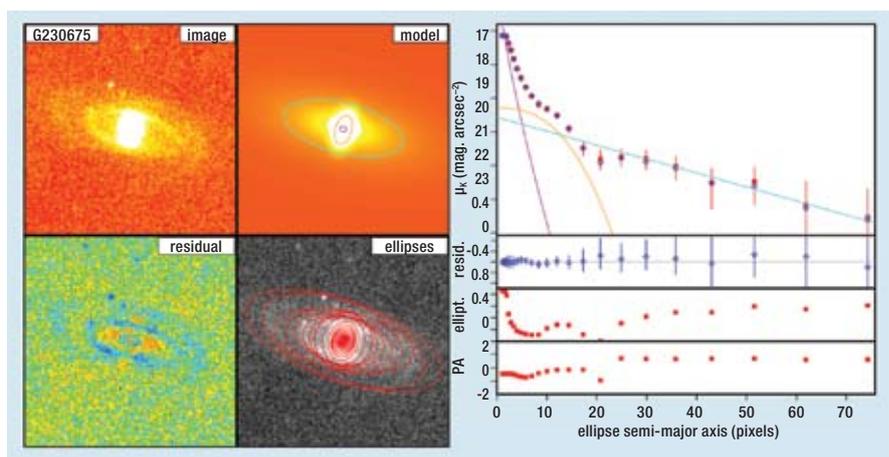

8: An overview of the information which will be extracted for GAMA galaxies. The upper panel shows the spectral information which we will extract for the entire dataset while the lower panel shows the structural information we will extract from the $z<0.1$ data. The galaxy shown in the lower panel is representative of the final data quality for a luminous spiral galaxy at $z=0.05$ where the bulge, bar and disc can be readily separated.

● **Mass measurements.** Initially this will consist of simple stellar mass estimates plus velocity dispersion measurements from GANDALF combined with effective radius measurements. Once ASKAP is operational this will be extended to include H$_I$ and Tully-Fisher dynamical mass estimates leading to total baryonic masses. The super massive black hole (SMBH) and dust masses will be provided by the structural and SED pipelines.

as moderate to deep imaging data from UV through IR to radio wavelengths. The reason high-resolution multiwavelength data is vital is the complex nature of the processes encountered by galaxies throughout their lifetime, including mergers, accretion, infalling cooling gas, active galactic nuclei, star formation, dynamical and chemical feedback, outflows and tidal interactions. As a result, galaxies comprise many distinct components (nucleus, bar, bulge, disc, halo) and phases of matter (hot gas, cold gas, stars, dust), and can look very different depending on the wavelength at which they are observed. This is demonstrated in figure 7 which shows NGC891 in the multiple wavelengths covered by GAMA. The FUV samples active star-forming regions, the optical wavelengths the younger stellar population, the near-IR the old stellar population and the mid-IR and far-IR the PAH and dust emission. Given this complexity, we contend that a full picture of galaxy evolution is unlikely to emerge from detailed studies of small (often unrepresentative) samples alone. Rather, one must build a comprehensive database that samples the entire parameter space over which galaxies exist. It is the broader aim of GAMA to provide such a dataset – with as much breadth, depth and detail as is required to cover all of the above galaxy-shaping processes, and to be able to separate galaxies into their components and constituents.

Our effort builds on the notion that databases are now as important as facilities, a concept perhaps best exemplified by the Sloan Digital Sky Survey (SDSS; more than 2000 papers with more than 50 000 citations in total: Strauss *et al.* 2009; Trimble *et al.* 2008). When complete, the GAMA project will in fact be significantly larger in terms of data volume than the entire SDSS (see table 3). However, a database is only as good as its instruments (in this case software instruments) and the manner in which the data are delivered. The four major software instruments currently planned for the GAMA data are described in box B.

### Engaging with the wider community

The study of galaxies is presently fragmented in a number of ways, mainly by wavelength (X-ray, optical, far-IR, sub-mm, radio) but also between those who study galaxies in exquisite detail (e.g. SINGS, THINGS) and those who study the coarse statistical properties of the population as a whole (e.g. SDSS, 2dFGRS). In addition there are high and low-redshift communities and those who study very specific, relatively rare galaxies (e.g. SCUBA sources, E+A galaxies). Often these subdisciplines do not interact as closely as one might like and have tended to evolve apart. The GAMA database has the potential to overcome at least some of these divisions by providing a comprehensive sample with detailed information, spanning wavelengths from UV to radio, a relatively large redshift range and utilizing a multitude of advanced analysis techniques (e.g. bulge-disc decomposition, spectral line analysis, spectral energy distribution modelling, and direct dynamical measurements). In due course, by bringing 8 m class telescopes to bear and with the advent of ALMA, JWST and SKA, the GAMA database has the potential to be extended in area, redshift, flux limit and detail of information. We therefore hope that the GAMA regions will develop into a natural starting point for new major survey programmes.

For the moment the GAMA survey has only just begun; nevertheless, we have already obtained more than 90 000 spectra in just 43 nights of allocated time in the first two years of AAT observations. Further rapid advances can be anticipated, with data concurrently being gathered by UKIRT, GALEX and GMRT; VISTA and Herschel observations should commence in the coming year, followed by VST, WISE and eventually ASKAP. Assembling this amount of data presents a tremendous logistical challenge, not least of which is the fundamentally different physical origins of the detected radiation. To complete the first stage of GAMA outlined here will take about five years, but the GAMA team is committed to issuing regular staged releases as and when complete data subsets become available.

Anyone wishing to become directly involved with the project or to start planning follow-on programmes is invited to contact the GAMA team via spd3@st-and.ac.uk. ●

### Table 3: GAMA and SDSS survey parameters

| parameter | GAMA | SDSS |
|---|---|---|
| galaxy redshifts | 275k | 700k |
| sky coverage (deg.$^2$) | 250 | ~8000 |
| spectral resolution (Å) | 4.6 | 3.3 |
| spectral range (Å) | 3700–8800 | 3900–9100 |
| spec. *r* limit (mag.) | 19.8 | 17.77 |
| $M^*$ z limit | 0.27 | 0.11 |
| $M^*$ volume (h$^{-3}$Mpc$^3$) | 6.6×10$^6$ | 25.9×10$^6$ |
| imaging bands | 21 | 5 |
| spatial resolution (″) | 0.7 | 1.5 |
| λ range (μm) | 0.15–10$^6$ | 0.3–0.9 |
| data volume | 120Tb–1Pb | 60Tb |

*Simon P Driver, School of Physics, University of St Andrews, North Haugh, St Andrews, KY16 9SS, UK (spd3@st-and.ac.uk); Peder Norberg, Ivan K Baldry, Steven P Bamford, Andrew M Hopkins, Jochen Liske, Jon Loveday, John A Peacock and the GAMA Team (listed below).*

*Acknowledgments. We are grateful for ongoing financial support from the Science and Technologies Facility Council and the Australian Research Council. SPD thanks the Australian Telescope National Facility and the University of Western Australia for financial support during the writing of this article. We are indebted to the staff of the Anglo-Australian Observatory for their dedication and professionalism in assisting with all aspects of the GAMA survey and look forward to continuing our tradition of Anglo-Australian collaboration beyond the UK's withdrawal from the AAO.*

## C: The UK bids farewell to the Anglo-Australian Observatory

On 1 June 2010 the UK formally withdraws from the Anglo-Australian Observatory (AAO), ending a 43-year partnership between the UK and Australian governments that led to the construction of the Anglo-Australian Telescope (AAT; first light April 1974), the operation of the UK Schmidt Telescope (UKST; since 1988), and the establishment of a world-leading instrumentation group specializing in fibre positioners. Through these facilities, and associated instruments, comprehensive breakthroughs have been made in almost all branches of astronomy (from exoplanets to cosmology).

One statistic worth highlighting is that the AAO is responsible for ~35% of all known redshifts (~2 million known of which ~700 000 were measured by AAO facilities). The AAO's share of the redshift market is rapidly growing with the advent of AAOmega and the ongoing WiggleZ (Blake *et al*. 2007) and GAMA surveys. In fact, by the time of the UK's withdrawal the AAO's market share will have risen to ~40%. Table 4 lists some of the more recent AAO redshift surveys and the number of redshifts measured. However one slices or dices it, the AAO has been a superb investment producing a remarkable science return. In a recent review of worldwide astronomical facilities (Trimble and Ceja 2008) the AAT is listed as the most successful 4 m facility worldwide, with more than double the citation rate of any other 4 m class telescope. In comparison to all optical telescopes (terrestrial or otherwise), the AAT comes in at fifth place in terms of both productivity and impact (see Director's Message in the August 2008 AAO Newsletter).

The partnership between the UK and Australian governments began in 1967 after 10 years of heated dialogue (Lovell *et al*. 1985) with an agreement to start construction of the AAT. Land was leased from the Mt Stromlo Observatory, which already operated a number of telescopes at the Siding Spring site, and construction of the AAT was completed in 1974 (Gascoigne *et al*. 1990). Since that time the AAT has seen a number of instruments come and go and is currently in the progress of building its next-generation instrument (HERMES), which will conduct a unique survey of galactic stars to unveil the sequence by which our home galaxy assembled itself. In the meantime the WiggleZ and GAMA surveys are set to continue the AAO's world-leading role in mapping the cosmos.

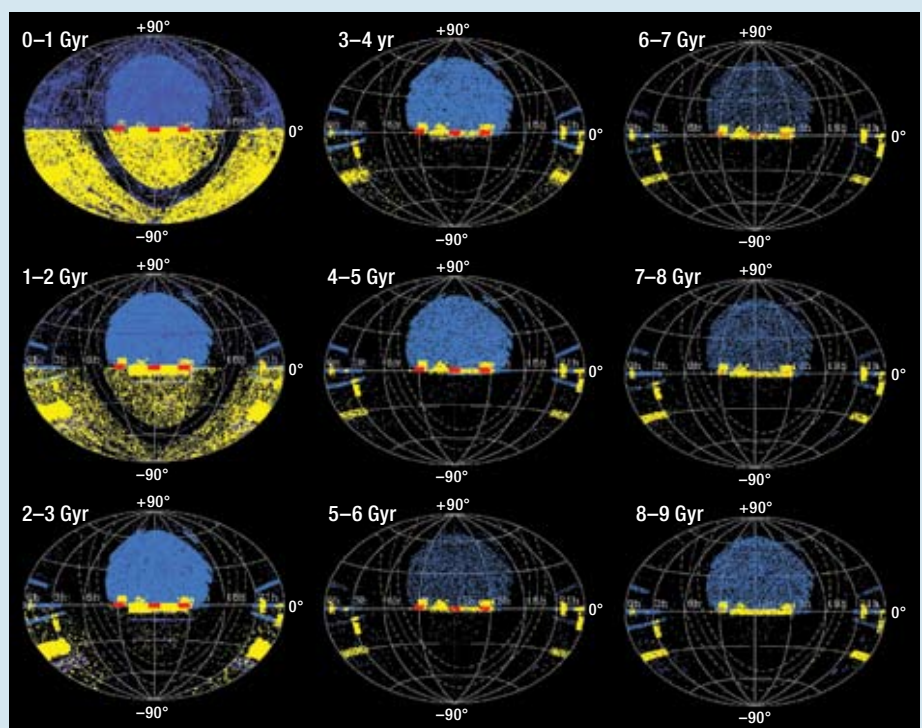

9: The distribution of all known redshifts on the sky (galaxies and QSOs) in intervals of 1 Gyr in lookback time from 0–9 Gyrs as indicated (0.0–1.38 in redshift). Data obtained by: the Anglo-Australian Observatory (yellow; 2dfGRS, 2QZ, 6dfGS, MGC, 2SLAQ) except for GAMA (red); and all other sources (blue; SDSS, CfA ZCAT, ESP, LCRS, zCOSMOS, VVDS, DEEP2).

### Table 4: Redshift surveys from Anglo-Australian Observatory

| survey | dates | observations |
|---|---|---|
| 2dFGRS | 1996–2001 | 227k |
| 2QZ | 1996–2001 | 28k |
| MGC | 2002–2006 | 8k |
| 2SLAQ-LRG | 2003–2006 | 16k |
| 2SLAQ-QSO | 2003–2006 | 3k |
| 6dFGS | 2003–2008 | 110k |
| UCD/Fornax | 2002–2006 | 16k |
| AUS | 2006+ | 50k |
| WiggleZ | 2007+ | 140k (200k) |
| GAMA | 2008+ | 90k (250k) |
| total | | 688k (910k) |

## The GAMA team

S P Driver, D T Hill, L S Kelvin, A S G Robotham, University of St Andrews, UK.
P Norberg, J A Peacock, N J G Cross, H R Parkinson, University of Edinburgh, UK.
I K Baldry, M Prescott, Liverpool John Moores University, UK.
S P Bamford, C J Conselice, L Dunne, University of Nottingham, UK.
A M Hopkins, S Brough, H Jones, R G Sharp, Anglo Australian Observatory, Australia.
J Liske, E van Kampen, European Southern Observatory, Germany.
J Loveday, S Oliver, I G Roseboom, University of Sussex, UK.
J Bland-Hawthorn, S M Croom, S Ellis, University of Sydney, Australia.
E Cameron, Swiss Federal Institute of Technology (ETH), Switzerland.
S Cole, C S Frenk, Durham University, UK.
W J Couch, Alister W Graham, R Proctor, Swinburne University of Technology, Australia.
R De Propris, Cerro Tololo Inter-American Observatory, Chile.
I F Doyle, E M Edmondson, R C Nichol, D Thomas, University of Portsmouth, UK.
S A Eales, Cardiff University, UK.
M J Jarvis, University of Hertfordshire, UK.
K Kuijken, University of Leiden, Netherlands.
O Lahav, University College London, UK.
B F Madore, M Seibert, Carnegie Institute of Science, USA.
M J Meyer, L Staveley-Smith, University of Western Australia, Australia.
S Phillipps, University of Bristol, UK.
C C Popescu, A E Sansom, University of Central Lancashire, UK.
W J Sutherland, Queen Mary University London, UK.
R J Tuffs, Max-Planck Institute for Nuclear Physics (MPIK), Germany.
S J Warren, Imperial College London, UK.